\documentclass[12pt,letterpaper]{article}
\usepackage{amsmath,amssymb,array,calc,rotating,epsfig,psfrag,amscd, cite}

\numberwithin{equation}{section}

\makeatletter
\renewcommand\section{\@startsection {section}{1}{\z@}%
                                   {-3.5ex \@plus -1ex \@minus -.2ex}
                                   {2.3ex \@plus.2ex}%
                                   {\normalfont\large\bfseries}}
\renewcommand\subsection{\@startsection{subsection}{2}{\z@}%
                                     {-3.25ex\@plus -1ex \@minus -.2ex}%
                                     {1.5ex \@plus .2ex}%
                                     {\normalfont\bfseries}}
\makeatother

\let\non\nonumber

\let\s=\sigma

\let\S=\Sigma

\newcommand{\bea}{\begin{eqnarray}}
\newcommand{\eea}{\end{eqnarray}}
\newcommand{\be}{\begin{equation}}
\newcommand{\ee}{\end{equation}}

\renewcommand{\O}{\operatorname{O}}

\newcommand{\G}{\Gamma}

\newcommand{\m}{\mu}

\newcommand{\p}{\partial}

\newcommand{\C}[1]{$(\ref{#1})$}

\def\IZ{\relax\ifmmode\mathchoice
{\hbox{\cmss Z\kern-.4em Z}}{\hbox{\cmss Z\kern-.4em Z}}
{\lower.9pt\hbox{\cmsss Z\kern-.4em Z}} {\lower1.2pt\hbox{\cmsss
Z\kern-.4em Z}}\else{\cmss Z\kern-.4em Z}\fi}
\def\IR{\relax{\rm I\kern-.18em R}}

\def\one{{\hbox{ 1\kern-.8mm l}}}

\newlength{\bredde}
\def\slash#1{\settowidth{\bredde}{$#1$}\ifmmode\,\raisebox{.15ex}{/}
\hspace*{-\bredde} #1\else$\,\raisebox{.15ex}{/}\hspace*{-\bredde}
#1$\fi}

\newsavebox{\zzzbar}
\sbox{\zzzbar}
  {\setlength{\unitlength}{0.9em}
  \begin{picture}(0.6,0.7)
  \thinlines
  \put(0,0){\line(1,0){0.6}}
  \put(0,0.75){\line(1,0){0.575}}
  \multiput(0,0)(0.0125,0.025){30}{\rule{0.3pt}{0.3pt}}
  \multiput(0.2,0)(0.0125,0.025){30}{\rule{0.3pt}{0.3pt}}
  \put(0,0.75){\line(0,-1){0.15}}
  \put(0.015,0.75){\line(0,-1){0.1}}
  \put(0.03,0.75){\line(0,-1){0.075}}
  \put(0.045,0.75){\line(0,-1){0.05}}
  \put(0.05,0.75){\line(0,-1){0.025}}
  \put(0.6,0){\line(0,1){0.15}}
  \put(0.585,0){\line(0,1){0.1}}
  \put(0.57,0){\line(0,1){0.075}}
  \put(0.555,0){\line(0,1){0.05}}
  \put(0.55,0){\line(0,1){0.025}}
  \end{picture}}

\newcommand{\ena}{\end{eqnarray}}
\newcommand{\beqa}{\begin{eqnarray}}
\newcommand{\eeqa}{\end{eqnarray}}

\def\G{\Gamma}

\newcommand{\g}{\gamma}

\def\g{\gamma}

\def\m{\mu}

\def\s{\sigma}

\def\G{\Gamma}

\def\L{\Lambda}
\def\O{\Omega}

\def\S{\Sigma}

\renewcommand{\O}{\operatorname{O}}

\begin{document}
\begin{titlepage}

\begin{center}



\vskip 2 cm
{\Large \bf Zero mode of the Fourier series of some modular graphs from Poincar\'e series}\\
\vskip 1.25 cm { Anirban Basu\footnote{email address:
    anirbanbasu@hri.res.in} } \\
{\vskip 0.5cm  Harish--Chandra Research Institute, HBNI, Chhatnag Road, Jhusi,\\
Prayagraj 211019, India}

\end{center}

\vskip 2 cm

\begin{abstract}
\baselineskip=18pt

We consider specific linear combinations of two loop modular graph functions on the toroidal worldsheet with $2s$ links for $s=2, 3$ and $4$. In each case, it satisfies an eigenvalue equation with source terms involving $E_{2s}$ and $E_s^2$ only. On removing certain combinations of $E_{2s}$ and $E_s^2$ from it, we express the resulting expression as an absolutely convergent Poincar\'e series.  This is used to calculate the power behaved terms in the asymptotic expansion of the zero mode of the Fourier expansion of these graphs in a simple manner.   

\end{abstract}

\end{titlepage}


\section{Introduction}

We consider modular graph functions~\cite{DHoker:2015gmr,DHoker:2015wxz} that arise  in the low momentum expansion of multi--graviton amplitudes in type II string theory at genus one. The amplitude is obtained by integrating products of them  over the truncated fundamental domain of $SL(2,\mathbb{Z})$ and keeping the finite contributions~\cite{Green:1999pv,Green:2008uj} which yields terms in the effective action that are analytic in the external momenta. These $SL(2,\mathbb{Z})$ invariant graphs have links that are given by the scalar Green function on the toroidal worldsheet, while the vertices are given by insertion points of vertex operators which are integrated over the worldsheet\footnote{More generally, the integrands of the string amplitudes at genus one involve $SL(2,\mathbb{Z})$ covariant modular graph forms which we shall not consider.}.

Hence it is useful to understand detailed properties of these modular graphs. Non--trivial graphs are one particle irreducible and also do not have any vertex with only one link ending on it, which follows from the properties of the Green function. These graphs satisfy eigenvalue equations on moduli space which have proved extremely useful in understanding their various properties as well as performing the integral over the truncated fundamental domain~\cite{Green:1999pv,Green:2008uj,DHoker:2015gmr,Basu:2015ayg,DHoker:2015wxz,Basu:2016fpd,DHoker:2016mwo,Basu:2016xrt,Basu:2016kli,DHoker:2016quv,Kleinschmidt:2017ege,Broedel:2018izr,DHoker:2019blr,Basu:2019idd,Gerken:2019cxz,Gerken:2020yii}.  

One loop modular graphs are given by the non-holomorphic Eisenstein series whose Fourier and Poincar\'e series are simple and well known. These graphs satisfy Laplace equation on moduli space. However, this analysis gets involved beyond one loop. Generically such graphs satisfy Poisson equations on moduli space, where the source terms involve modular graphs such that every term in the eigenvalue equation preserves the number of links. At two loops, the eigenvalue equation satisfied by the graphs is known~\cite{DHoker:2015gmr,Basu:2019idd}, while explicit expressions for the Fourier and Poincar\'e series are also known~\cite{DHoker:2015gmr,DHoker:2017zhq,Ahlen:2018wng,DHoker:2019txf,Dorigoni:2019yoq,DHoker:2019mib}. The eigenvalue equation has been derived either by manipulating the expression for the graphs given as lattice sums after acting on them with the Laplacian operator, or by analyzing the variations of the graphs on varying the Beltrami differentials. The Fourier and Poincar\'e series have been derived both using the lattice sum representations of these graphs, and the differential equations they satisfy.  

In this paper, we reconsider these issues for some two loop modular graphs from a different viewpoint. If the graph has the Fourier series given by
\be F(\tau) = \sum_{n\in\mathbb{Z}}F_n(\tau_2)e^{2\pi i n\tau_1},\ee
where $\tau = \tau_1+i\tau_2$ is the complex structure of the torus, the zero mode of the Fourier series is simply given by $F_0(\tau_2)$ (which is also referred to as the constant term). This (as well as every other $F_n(\tau_2)$) has terms which are power behaved in $\tau_2$ as well as terms that are exponentially suppressed in $\tau_2$ in the large $\tau_2$ expansion. We want to derive the terms that are power behaved in $\tau_2$ not by directly solving the eigenvalue equation it satisfies but from the Poincar\'e series the graph satisfies. Thus in general while it is interesting to obtain the Fourier series directly from the Poincar\'e series, this is the simplest exercise one can think of carrying out. Though here we shall reproduce known results for the Fourier series, the method is general enough to be possibly extended to other cases.   

We now briefly outline the strategy we follow. We consider (linear combinations of) modular graphs $\Psi_{2s}$ with $2s$ links for $s=2,3$ and $4$. Each of them satisfies an eigenvalue equation with source terms that involve only $E_{2s}$ and $E_s^2$, where $E_p$ is the non--holomorphic Eisenstein series. As stressed above, we want to understand the Fourier modes from the Poincar\'e series rather than from the solutions of the eigenvalue equation that $\Psi_{2s}$ satisfies. In order to do so, we first obtain an eigenvalue equation the seed function of the Poincar\'e series for $\Psi_{2s}$ satisfies. We use a specific choice of boundary conditions to solve for such equations. If the source term in the eigenvalue equation of the seed function is $O(\tau_2^s)$ for large $\tau_2$, we impose the boundary condition that the constant term in the Fourier expansion of the seed function is $O(\tau_2^s)$ for large $\tau_2$, and $O(\tau_2^{1-s})$ for small $\tau_2$~\footnote{For $s=1$, the behavior is $O({\rm ln}\tau_2)$ for small $\tau_2$.}. The remaining Fourier modes are exponentially suppressed at large $\tau_2$ and are $O(\tau_2^{1-s})$ for small $\tau_2$. This choice of boundary conditions yields a unique solution for the seed function.       

For the cases we consider, the seed function contains a term that is linear in $\tau_2$ and hence the Poincar\'e series is not absolutely convergent. To remedy this, we remove certain combinations of $E_{2s}$ and $E_s^2$ from $\Psi_{2s}$ such that the resulting expression is given by an absolutely convergent Poincar\'e series. Finally this is used to obtain the terms that are power behaved in the large $\tau_2$ expansion of the zero mode $F_0(\tau_2)$ of the Fourier series of the modular graph. Part of this work generalizes some of the results in~\cite{Ahlen:2018wng,Dorigoni:2019yoq} and offers a somewhat different viewpoint. We end with a discussion on possibly generalizing our analysis to a class of modular graphs and also discuss certain limitations.   
  
\section{Two loop modular graph functions on the toroidal worldsheet}     

We now define the two loop modular graphs. To begin with, note that the coordinate on the toroidal worldsheet $\S$ is given by
\be -\frac{1}{2} \leq {\rm Re} z \leq \frac{1}{2} , \quad 0 \leq {\rm Im} z \leq \tau_2.\ee 
The Green function representing the link connecting the vertices of the graph on the toroidal worldsheet is given by~\cite{Lerche:1987qk,Green:1999pv}

\be G(z_i,z_j) = \frac{1}{\pi} \sum_{(m,n)\neq (0,0)} \frac{\tau_2}{\vert m\tau+n\vert^2}e^{\pi[\overline{z}_{ij}(m\tau+n)-z_{ij}(m\overline{\tau}+n)]/\tau_2}\ee
where $z_{ij} = z_i - z_j$.

We now define\footnote{Here $\S^a$ means $a$ copies of $\S$.} the product of $a$ adjoining links by
\be {\mathcal{G}} (z_1,z_{a+1};a) \equiv \int_{\S^{a-1}}\prod_{i=2}^a \frac{d^2 z_i}{\tau_2} G(z_1,z_2) G(z_2,z_3) \ldots G(z_a,z_{a+1}),\ee
where the first and the last insertion points are not integrated over.

Then the two loop modular graph $C_{a,b,c}$ which is symmetric under interchange of $a,b$ and $c$, is given by
\be \label{defCabc} C_{a,b,c} = \int_{\S^2} \prod_{i=1}^2 \frac{d^2 z_i}{\tau_2} \mathcal{G} (z_1,z_2;a) \mathcal{G} (z_1,z_2;b) \mathcal{G} (z_1,z_2;c).\ee

In this paper, we shall be interested in some two loop modular graphs with even number of links. The simplest graph $C_{1,1,2}$ which we consider has four links. With six links, we shall consider the graphs $C_{1,2,3}$ and $C_{2,2,2}$. Finally, with eight links, we shall consider a specific linear combination of the graphs $C_{1,3,4}$, $C_{2,2,4}$ and $C_{2,3,3}$. In every case, the eigenvalue equation has source terms that only involve $E_{2s}$ and $E_s^2$ and no other graphs. The reason for this choice will be apparent later.

\section{The analysis of the modular graph $C_{1,1,2}$ with four links}

\subsection{The absolutely convergent Poincar\'e series}

The modular graph $C_{1,1,2}$ satisfies the eigenvalue equation~\cite{DHoker:2015gmr,Basu:2016kli}
\be \Big(\Delta -2\Big) C_{1,1,2} = 9 E_4 - E_2^2,\ee
where the $SL(2,\mathbb{Z})$ invariant Laplacian is defined by
\be \Delta = 4\tau_2^2\frac{\p^2}{\p\tau \p\bar\tau},\ee
and the non--holomorphic Eisenstein series that arise as the source terms is defined in appendix A. This leads to the Poincar\'e series representation~\cite{DHoker:2015gmr}
\be \label{main1}C_{1,1,2} -\frac{2}{3}E_4 = \sum_{\g \in \G_\infty \backslash SL(2,\mathbb{Z})} \Lambda_{1,1,2} (\g(\tau)),\ee
where the seed function is given by
\be \label{P1}\Lambda_{1,1,2} (\tau) = \frac{\pi \zeta(3)}{90}\tau_2 +\frac{\pi \tau_2}{90} \sum_{n=1}^\infty \frac{\s_3(n)}{n^3} \Big(e^{2\pi i n \tau}+ e^{-2\pi i n \overline\tau}\Big).\ee
Note that the Poincar\'e series is not for the modular graph $C_{1,1,2}$ but a modular invariant combination involving a shift by $-2E_4/3$. 

The Poincar\'e series in \C{P1} is not absolutely convergent~\footnote{Issues regarding absolute convergence of Poincar\'e series that arise in integrals at genus one in string theory have been considered in a different context in~\cite{Angelantonj:2011br,Angelantonj:2012gw}.} because of the term linear in $\tau_2$ in $\L_{1,1,2}$~\footnote{This is exactly for the same reason that the Poincar\'e series \C{P} is absolutely convergent only for ${\rm Re}s>1$. In fact, the case for $s=1$ needs to be regularized and the $\tau_1$ independent contributions have $\tau_2$ dependence given by $\tau_2$ and ${\rm ln}\tau_2$.}. This is problematic for our purposes of calculating the zero mode of the Fourier series as we shall soon see, as the calculation involves performing an integral involving each individual term in the seed function. To remedy this, we consider
\be \label{con1}E_2^2 -\frac{7}{3}E_4\ee
which using \C{P}, yields the Poincar\'e series
\be E_2^2 -\frac{7}{3}E_4 = \sum_{\g \in \G_\infty \backslash SL(2,\mathbb{Z})} \Upsilon_1 (\g(\tau)),\ee
where the seed function is given by
\be \Upsilon_1 (\tau) = 2\Lambda_{1,1,2}  + \frac{4\zeta(4)\tau_2^2}{\pi^2} \sum_{n=1}^\infty \frac{\s_3(n)}{n^2} \Big(e^{2\pi i n \tau}+ e^{-2\pi i n \overline\tau}\Big).\ee
Thus the combination in \C{con1} is precisely such that its seed function includes the problematic term that appears in the seed function in \C{main1} which is an obstruction to the absolute convergence of the Poincar\'e series.  

Using this, we have that
\be \label{O112}C_{1,1,2} + \frac{1}{2} \Big(E_4 - E_2^2\Big) = \sum_{\g \in \G_\infty \backslash SL(2,\mathbb{Z})} \Omega_{1,1,2}(\g(\tau)),\ee
where the seed function is given by
\be \Omega_{1,1,2} (\tau) =  -\frac{2\zeta(4)\tau_2^2}{\pi^2} \sum_{n=1}^\infty \frac{\s_3(n)}{n^2} \Big(e^{2\pi i n \tau}+ e^{-2\pi i n \overline\tau}\Big).\ee
Hence we see that $C_{1,1,2} +(E_4- E_2^2)/2$ is given by a Poincar\'e series that is absolutely convergent. In fact each term in the Fourier expansion of the seed function is exponentially suppressed at the cusp. We now calculate the zero mode of the Fourier series of $C_{1,1,2}$.

\subsection{Zero mode of the Fourier series}

To calculate the zero mode of the Fourier series of $C_{1,1,2}$, we use the results given in appendix B. Thus in \C{Fourier}, substituting 
\be \label{add1}F = C_{1,1,2} + \frac{1}{2} \Big(E_4 - E_2^2\Big),\ee 
from \C{O112} and \C{Seed}, we have that
\be \Omega_0 = 0, \quad \Omega_m = -\frac{2\zeta(4)\tau_2^2\s_3(m)}{\pi^2 m^2} e^{-2\pi \vert m \vert \tau_2} ~(m \neq 0).\ee
This yields the expression for the zero mode on using \C{Fourier1}, leading to
\be \label{exp}F_0 = -\frac{2\zeta(4)}{\pi^2\tau_2}\sum_{n>0}\sum_{m\neq 0} \frac{S(m,0;n)\s_3(m)}{m^2 n^4}\int_{-\infty}^\infty \frac{dx}{(1+x^2)^2}e^{-2\pi(\vert m \vert +imx)/[n^2\tau_2(1+x^2)]}.\ee
We now analyze this expression along the lines of~\cite{Ahlen:2018wng}. We first expand the exponential in an infinite series and perform the sum over $n$ using \C{Fourier2}. We next perform the $x$ integral using the relation
\be \int_{-\infty}^\infty \frac{dx}{(1+ix)^a(1-ix)^b} = 2^{2-(a+b)}\pi\frac{\G(a+b-1)}{\G(a)\G(b)},\ee
for ${\rm Re}(a+b)>1$. 
This gives us that
\be \label{changesum}F_0 = -\frac{\zeta(4)}{\pi\tau_2} \sum_{k=0}^\infty \frac{(k+2)}{k! \zeta(2k+4)}\Big(-\frac{\pi}{\tau_2}\Big)^k\sum_{m>0} \frac{\s_3(m)\s_{-3-2k}(m)}{m^{2-k}}.\ee
In this expression, the sum over $k$ arises from expanding the exponential in the integrand in \C{exp}, and we have interchanged the sums over $m$ and $k$~\footnote{The sum over $k$ is formally divergent since it grows as $m^k$ for fixed $m$, and hence interchanging the two sums is actually not allowed. However, we still proceed with the calculation and see what it gives us.}. We now perform the sum over $m$ using the identity  
\be \label{divsum}\sum_{m>0} \frac{\s_p(m)\s_q(m)}{m^r} = \frac{\zeta(r)\zeta(r-p)\zeta(r-q)\zeta(r-p-q)}{\zeta(2r-p-q)}\ee
which we analytically continue to all integral values of $r$. 
This gives us
\be F_0 = -\frac{1}{\pi\tau_2}\sum_{k=0}^\infty \frac{(k+2)\zeta(-1-k)\zeta(2-k)\zeta(2+k)\zeta(5+k)}{k!\zeta(4+2k)}\Big(-\frac{\pi}{\tau_2}\Big)^k.\ee
In this infinite sum only the $k=0,1$ and 2 terms contribute, while the rest vanish. This gives us
\be F_0 = \frac{5\zeta(5)}{12\pi\tau_2} - \frac{3\zeta(3)^2}{4\pi^2\tau_2^2} +\frac{7\zeta(7)}{8\pi^3\tau_2^3},\ee
where the $k=1$ contribution has to be regularized\footnote{This is done using
\be \zeta(-1-k) = \frac{\zeta(2+k)\G(1+k/2)}{\pi^{k+3/2}\G(-1/2-k/2)},\ee
which leads to (as $k\rightarrow 1$)
\be \zeta(-1-k)\zeta(2-k)= -\frac{\zeta(3)\zeta(1+\epsilon)}{2\pi^2\G(\epsilon/2)} \rightarrow -\frac{\zeta(3)}{4\pi^2}\ee
as $\epsilon = 1-k \rightarrow 0$.}. 

Thus having obtained the power behaved terms in the zero mode in the Fourier expansion of \C{add1} it is straightforward to obtain those in the Fourier expansion of $C_{1,1,2}$ using \C{Eisen}. This gives us
\bea C_{1,1,2} = \frac{4\zeta(8)\tau_2^4}{3\pi^4}+\frac{\zeta(3)\pi\tau_2}{45} + \frac{5\zeta(5)}{12\pi\tau_2} - \frac{\zeta(3)^2}{4\pi^2\tau_2^2} +\frac{9\zeta(7)}{16\pi^3\tau_2^3}\eea
which precisely agrees with the expression in~\cite{DHoker:2015gmr}.

\section{The analysis of the modular graphs $C_{1,2,3}$ and $C_{2,2,2}$ with six links}

\subsection{The absolutely convergent Poincar\'e series}

The modular graphs $C_{1,2,3}$ and $C_{2,2,2}$ satisfy the coupled eigenvalue equations~\cite{DHoker:2015gmr,Basu:2016kli}
\bea \label{coupled}\Big(\Delta -2\Big) \Big(4C_{1,2,3} + C_{2,2,2}\Big) &=& 52 E_6 - 4 E_3^2, \non \\ \Big(\Delta -12\Big)\Big(6 C_{1,2,3} - C_{2,2,2}\Big) &=& 108 E_6 - 36 E_3^2.\eea
We now obtain absolutely convergent Poincar\'e series for modular invariant expressions involving these graphs.  

Using the relation \C{Eisenval} we rewrite the first equation in \C{coupled} as
\be \label{rhs1}(\Delta -2) \Psi_1 = \frac{5720}{691}E_6 - 4 E_3^2,\ee 
where 
\be \label{Psi1}\Psi_1 = 4 C_{1,2,3} + C_{2,2,2} - \frac{1079}{691}E_6.\ee
The motivation for writing it in this way will be clear shortly.  

We now want to obtain the Poincar\'e series for $\Psi_1$. Letting
\be \Psi_1 = \sum_{\g \in \G_\infty \backslash SL(2,\mathbb{Z})} \Phi_1 (\g(\tau)),\ee
we see that the seed function $\Phi_1$ satisfies the eigenvalue equation
\be \label{rhs2}(\Delta -2) \Phi_1 = \frac{16\pi^6}{893025} \tau_2^6 - \frac{8\pi^3}{945}\tau_2^3 E_3\ee
where we have used \C{P}. Now from \C{Eisen} we see that the right hand side of \C{rhs2} is $O(\tau_2)$ for large $\tau_2$. We shall now use this fact to choose boundary conditions to uniquely solve for the seed function, thus justifying the choice of \C{Psi1}. 

To understand the choice of boundary conditions we take to uniquely solve for the seed function $\Phi(\tau)$, consider the generic eigenvalue equation\footnote{For all the cases we consider, $\lambda = \m(\m-1)$ where $\m$ is a non--zero integer.}
\be \Big(\Delta - \lambda \Big) \Phi = f(\tau)\ee
it satisfies. Suppose $f(\tau)$ is $O(\tau_2^s)$ for large $\tau_2$. Let the Fourier series of $\Phi$ be given by
\be \Phi (\tau) =  \sum_{n \in \mathbb{Z}} \Phi_{n}(\tau_2) e^{2\pi i n \tau_1}.\ee  
Then for $\Phi_0(\tau_2)$, we choose the boundary condition that it is $O(\tau_2^s)$ for large $\tau_2$, and it is $O(\tau_2^{1-s})$ for small $\tau_2$. For $\Phi_n (\tau_2)$ $(n \neq 0)$ we choose the boundary condition that it is $O(\tau_2^{1-s})$ for small $\tau_2$~\footnote{This mode is exponentially suppressed for large $\tau_2$, and hence we do not need to impose any boundary condition at large $\tau_2$ to solve for it.}. This non--trivial choice of boundary conditions is motivated by~\cite{Green:2014yxa}. However, the analysis there is done for a modular invariant eigenfunction, which is not the case for the seed function\footnote{In fact, for $s=1$, the analysis of~\cite{Green:2014yxa} leads to $O({\rm ln}\tau_2)$ behavior for small $\tau_2$, rather than $\tau_2^0$.}. Hence we simply take it as the definition of the choice of boundary conditions for calculating the seed function. We shall see this yields the correct answer for the Fourier mode of the modular graph in all the cases we consider (similar issues have also been discussed in~\cite{Ahlen:2018wng}). It would be interesting to understand the issue of boundary conditions to be imposed on the Fourier modes of the seed function in detail.  
 
Thus for the case at hand, we impose the boundary condition that the zero mode of the Fourier series of the seed function is $O(\tau_2)$ for large $\tau_2$, while all the modes are $O({\rm ln}\tau_2)$ for small $\tau_2$ (whether it is $\O({\rm ln}\tau_2)$ or $O(\tau_2^0)$ for small $\tau_2$ will be irrelevant for our purposes). In fact, this will be the boundary conditions we shall impose for all the cases we consider because $f(\tau)$ is always $O(\tau_2)$ for large $\tau_2$, as we shall see. 

Thus setting
\be \Phi_1 (\tau) = \Phi_{1,0} (\tau_2) + \sum_{n \neq 0} \Phi_{1,n}(\tau_2) e^{2\pi i n \tau_1},\ee
we see that the zero mode $\Phi_{1,0} (\tau_2)$ satisfies
\be \Big(\tau_2^2\frac{d^2}{d\tau_2^2} -2\Big) \Phi_{1,0} = -\frac{2\pi\zeta(5)}{315}\tau_2,\ee
which is solved by
\be \Phi_{1,0} (\tau_2)= \frac{\pi \zeta(5)}{315}\tau_2\ee
on using the boundary conditions at large and small $\tau_2$ which remove the homogeneous solution.

The non--zero mode $\Phi_{1,n} (n \neq 0)$ satisfies the equation
\bea \Big(\tau_2^2\frac{d^2}{d\tau_2^2} -2 - 4\pi^2 n^2\tau_2^2\Big)\Phi_{1,n} = -\frac{8\pi^3\tau_2^3 \s_5(n)}{945\vert n \vert^3}e^{-2\pi \vert n \vert \tau_2 } \Big(1+\frac{3}{2\pi \vert n \vert \tau_2}+\frac{3}{4\pi^2n^2\tau_2^2}\Big).\eea

We express the solution as
\be \Phi_{1,n} = \Phi_{1,n}^h +\Phi_{1,n}^p,\ee
where $\Phi_{1,n}^h$ and $\Phi_{1,n}^p$ are the solutions to the homogeneous equation and the particular solution respectively. 
While the solution to the homogeneous equation is given by\footnote{Here and in the cases to be considered later where the analysis is similar, we keep the solution involving $\sqrt{\tau_2}K_s (2\pi\vert n \vert\tau_2)$ and ignore the linearly independent solution involving $\sqrt{\tau_2} I_s(2\pi\vert n \vert \tau_2)$ on physical grounds, as it diverges exponentially for large $\tau_2$.}
\be \Phi_{1,n}^h = a_{1,n} \sqrt{\tau_2} K_{3/2}(2\pi \vert n \vert\tau_2),\ee
where $a_{1,n}$ is an arbitrary constant, the particular solution is given by
\bea \Phi_{1,n}^p = \frac{\s_5 (n)}{1890n^6}\Big( \frac{15}{8\pi \vert n \vert \tau_2} +\frac{15}{4} + 6\pi \vert n \vert \tau_2 + 2\pi^2 n^2 \tau_2^2\Big) e^{-2\pi \vert n \vert \tau_2}.\eea
Thus the boundary condition at small $\tau_2$ which demands cancellation of the $O(\tau_2^{-1})$ contribution yields
\be a_{1,n} = -\frac{\s_5 (n)}{252 \vert n \vert^{11/2}},\ee
leading to the solution
\be \Phi_{1,n} (\tau_2) = \frac{\pi\s_5 (n)\tau_2}{945 \vert n\vert^5} \Big(3+ \pi \vert n \vert \tau_2\Big)e^{-2\pi \vert n \vert \tau_2}.\ee

Thus we see that
\be \label{123} 4C_{1,2,3} + C_{2,2,2} - \frac{1079}{691}E_6= \sum_{\g \in \G_\infty \backslash SL(2,\mathbb{Z})} \L_1 (\g(\tau)),\ee
where the seed function is given by
\be \label{123seed}\L_1 (\tau) =  \frac{\pi\zeta(5)}{315}\tau_2 + \frac{\pi \tau_2}{945} \sum_{n=1}^\infty \frac{\s_5(n)}{n^5} \Big(3+ \pi n \tau_2\Big)\Big(e^{2\pi i n \tau}+ e^{-2\pi i n \bar\tau}\Big).\ee

Proceeding similarly, we rewrite the second equation in \C{coupled} as
\be (\Delta -12) \Psi_2 = \frac{51480}{691}E_6 - 36 E_3^2,\ee
where 
\be \Psi_2 = 6 C_{1,2,3} - C_{2,2,2} - \frac{1286}{691}E_6.\ee
The Poincar\'e series for $\Psi_2$ is given by
\be \Psi_2 = \sum_{\g \in \G_\infty \backslash SL(2,\mathbb{Z})} \Phi_2 (\g(\tau)),\ee
where the seed function satisfies
\be \label{rhs3}(\Delta -12) \Phi_2 = \frac{16\pi^6}{99225} \tau_2^6 - \frac{8\pi^3}{105}\tau_2^3 E_3.\ee
Once again, the right hand side of \C{rhs3} is $O(\tau_2)$ for large $\tau_2$.

Thus setting
\be \Phi_2 (\tau) = \Phi_{2,0} (\tau_2) + \sum_{n \neq 0} \Phi_{2,n}(\tau_2) e^{2\pi i n \tau_1},\ee
we see that $\Phi_{2,0} (\tau_2)$ satisfies
\be \Big(\tau_2^2\frac{d^2}{d\tau_2^2} -12\Big) \Phi_{2,0} = -\frac{2\pi\zeta(5)}{35}\tau_2,\ee
which is solved by
\be \Phi_{2,0} (\tau_2)= \frac{\pi \zeta(5)}{210}\tau_2.\ee

The non--zero mode $\Phi_{2,n} (n \neq 0)$ satisfies
\bea \Big(\tau_2^2\frac{d^2}{d\tau_2^2} -12 - 4\pi^2 n^2\tau_2^2\Big)\Phi_{2,n} = -\frac{8\pi^3\tau_2^3 \s_5(n)}{105\vert n \vert^3}e^{-2\pi \vert n \vert \tau_2 } \Big(1+\frac{3}{2\pi \vert n \vert \tau_2}+\frac{3}{4\pi^2n^2\tau_2^2}\Big).\eea
Once again, expressing the solution as 
\be \Phi_{2,n} = \Phi_{2,n}^h +\Phi_{2,n}^p,\ee
we see that the homogeneous equation is solved by
\be \Phi_{2,n}^h = a_{2,n} \sqrt{\tau_2} K_{7/2}(2\pi \vert n \vert\tau_2)\ee
for an arbitrary constant $a_{2,n}$, while the particular solution is given by
\bea \Phi_{2,n}^p = -\frac{\s_5 (n)}{840n^6}\Big( \frac{525}{8\pi^3 \vert n \vert^3 \tau_2^3} +\frac{525}{4\pi^2 n^2\tau_2^2} +\frac{105}{\pi \vert n \vert \tau_2}+ 35- 4 \pi \vert n \vert  \tau_2 - 8\pi^2 n^2 \tau_2^2\Big) e^{-2\pi \vert n \vert \tau_2}.\non \\ \eea

The boundary condition at small $\tau_2$ leads to the cancellation of $\tau_2^{-3}$ and $\tau_2^{-1}$ terms and yields
\be a_{2,n} = \frac{\s_5 (n)}{12 \vert n \vert^{11/2}},\ee
leading to the expression
\be \Phi_{2,n} (\tau_2) = \frac{\pi\s_5 (n)\tau_2}{210 \vert n\vert^5} \Big(1+ 2\pi \vert n \vert\tau_2\Big)e^{-2\pi \vert n \vert \tau_2}\ee
for the non--zero mode. 

This leads to
\be \label{222} 6C_{1,2,3} - C_{2,2,2} - \frac{1286}{691}E_6 = \sum_{\g \in \G_\infty \backslash SL(2,\mathbb{Z})} \L_2 (\g(\tau)),\ee
where the seed function is
\be \label{222seed}\L_2 (\tau) =  \frac{\pi\zeta(5)}{210}\tau_2 + \frac{\pi \tau_2}{210} \sum_{n=1}^\infty \frac{\s_5(n)}{n^5} \Big(1+ 2\pi n \tau_2\Big)\Big(e^{2\pi i n \tau}+ e^{-2\pi i n \bar\tau}\Big).\ee

Now the Poincar\'e series in \C{123} and \C{222} are not absolutely convergent because of the term linear in $\tau_2$ in $\L_{i}$ ($i=1,2$). To remedy this, we consider
\be E_3^2 -\frac{1430}{691}E_6 = \sum_{\g \in \G_\infty \backslash SL(2,\mathbb{Z})} \Upsilon_2 (\g(\tau)),\ee
where the seed function is given by
\be \label{defU}\Upsilon_2 (\tau) = \Lambda_{U} (\tau) + \frac{2\zeta(6)\tau_2^3}{\pi^3} \sum_{n=1}^\infty \frac{\s_5(n)}{n^3} \Big(1+\frac{3}{2\pi n \tau_2}\Big)\Big(e^{2\pi i n \tau}+ e^{-2\pi i n \overline\tau}\Big).\ee
In \C{defU}, the expression for $\L_U (\tau)$ is given by
\be \Lambda_U (\tau)= \frac{\pi\zeta(5)}{630}\tau_2 + \frac{\pi\tau_2}{630}\sum_{n=1}^\infty \frac{\s_5(n)}{n^5} \Big(e^{2\pi i n \tau}+ e^{-2\pi i n \overline\tau}\Big), \ee
which appears in \C{123seed} as well as in \C{222seed}. In fact, it is proportional to the problematic terms that arise in both $\Lambda_1$ and $\Lambda_2$ which are an obstruction to absolute convergence of the Poincar\'e series.   

Hence this immediately gives us that 
\bea \label{decoup}4C_{1,2,3} + C_{2,2,2} +\frac{1781}{691}E_6 - 2 E_3^2 &=&  \sum_{\g \in \G_\infty \backslash SL(2,\mathbb{Z})}  \chi_1 (\g(\tau)), \non \\ 6C_{1,2,3} - C_{2,2,2} +\frac{3004}{691}E_6 - 3 E_3^2 &=&  \sum_{\g \in \G_\infty \backslash SL(2,\mathbb{Z})}  \chi_2 (\g(\tau)),\eea
with the seed functions
\bea \chi_1 (\tau) &=& -\frac{4\zeta(6)\tau_2^3}{\pi^3}\sum_{n=1}^\infty \frac{\s_5(n)}{n^3} \Big(1+\frac{5}{4\pi n \tau_2}\Big)\Big(e^{2\pi i n \tau}+ e^{-2\pi i n \overline\tau}\Big), \non \\ \chi_2 (\tau) &= &-\frac{6\zeta(6)\tau_2^3}{\pi^3}\sum_{n=1}^\infty \frac{\s_5(n)}{n^3} \Big(e^{2\pi i n \tau}+ e^{-2\pi i n \overline\tau}\Big).\eea
Thus in \C{decoup} we obtain expressions for the modular invariant quantities on the left hand side in terms of absolutely convergent Poincar\'e series, in which every term in the Fourier expansion of the seed function is exponentially suppressed at the cusp. 

This allows us to obtain expressions involving the graphs $C_{1,2,3}$ and $C_{2,2,2}$ separately which are given by  
\bea \label{dec}C_{1,2,3} +\frac{957}{1382} E_6 - \frac{1}{2} E_3^2 &=&  \sum_{\g \in \G_\infty \backslash SL(2,\mathbb{Z})}  \Omega_{1,2,3} (\g(\tau)), \non \\ C_{2,2,2} - \frac{133}{691}E_6 &=&  \sum_{\g \in \G_\infty \backslash SL(2,\mathbb{Z})}  \Omega_{2,2,2} (\g(\tau)),\eea
where the seed functions are given by
\bea \label{dec2}\Omega_{1,2,3} (\tau) &=& -\frac{\zeta(6)\tau_2^3}{\pi^3}\sum_{n=1}^\infty \frac{\s_5(n)}{n^3} \Big(1+\frac{1}{2\pi n \tau_2}\Big)\Big(e^{2\pi i n \tau}+ e^{-2\pi i n \overline\tau}\Big), \non \\ \Omega_{2,2,2} (\tau) &=& -\frac{3\zeta(6)\tau_2^2}{\pi^4}\sum_{n=1}^\infty \frac{\s_5(n)}{n^4} \Big(e^{2\pi i n \tau}+ e^{-2\pi i n \overline\tau}\Big).\eea

\subsection{Zero mode of the Fourier series}

We now calculate the zero mode in the Fourier series expansion of the graphs $C_{1,2,3}$ and $C_{2,2,2}$ using results in appendix B. Since the steps involved in the analysis are the same as discussed previously, we skip some intermediate steps. 

We first consider $C_{2,2,2}$. In \C{Fourier}, for
\be F = C_{2,2,2} - \frac{133}{691}E_6,\ee
in \C{Seed} we have that 
\be \Omega_0 =0, \quad \Omega_m = -\frac{3\zeta(6)\tau_2^2\s_5(m)}{\pi^4 m^4}e^{-2\pi \vert m \vert \tau_2}~(m\neq 0).\ee
This gives us the expression
\bea  F_0 &=& -\frac{3\zeta(6)}{2\pi^3\tau_2} \sum_{k=0}^\infty \frac{(k+2)}{k! \zeta(2k+4)}\Big(-\frac{\pi}{\tau_2}\Big)^k\sum_{m>0} \frac{\s_5(m)\s_{-3-2k}(m)}{m^{4-k}} \non \\ &=&-\frac{3}{2\pi^3\tau_2}\sum_{k=0}^\infty \frac{(k+2)\zeta(-1-k)\zeta(4-k)\zeta(2+k)\zeta(7+k)}{k!\zeta(4+2k)}\Big(-\frac{\pi}{\tau_2}\Big)^k.\eea
Here only $k=0,2,3$ and $4$ contributes in the infinite sum, leading to
\be F_0 = \frac{\zeta(7)}{24\pi\tau_2} - \frac{7\zeta(9)}{16\pi^3\tau_2^3} +\frac{15\zeta(5)^2}{16\pi^4\tau_2^4}- \frac{32175\zeta(11)}{44224\pi^5\tau_2^5},\ee
where the $k=3$ term has to be regularized. This leads to
\be \label{asymp222}C_{2,2,2} = \frac{266\zeta(12)\tau_2^6}{691\pi^6} + \frac{\zeta(7)}{24\pi\tau_2} -\frac{7\zeta(9)}{16\pi^3\tau_2^3} +\frac{15\zeta(5)^2}{16\pi^4\tau_2^4} -\frac{81\zeta(11)}{128\pi^5\tau_2^5} \ee
which exactly agrees with the known result~\cite{DHoker:2016quv}.

Now from \C{dec}, \C{dec2} and \C{Fourier}, for
\be F = C_{1,2,3} - \frac{1}{6}C_{2,2,2} +\frac{1502}{2073}E_6 - \frac{1}{2}E_3^2,\ee
from \C{Seed}, we have that
\be \Omega_0 = 0, \quad \Omega_m = -\frac{\zeta(6)\tau_2^3\s_5(m)}{\pi^3 \vert m\vert^3}e^{-2\pi \vert m \vert \tau_2}~(m \neq 0).\ee
This leads to
\bea F_0 &=& -\frac{\zeta(6)}{16\pi^2\tau_2^2} \sum_{k=0}^\infty \frac{(k+3)(k+4)}{k!\zeta(2k+6)} \Big(-\frac{\pi}{\tau_2}\Big)^k \sum_{m>0}\frac{\s_5(m)\s_{-5-2k}(m)}{m^{3-k}} \non \\ &=&-\frac{1}{16\pi^2\tau_2^2}\sum_{k=0}^\infty \frac{(k+3)(k+4)\zeta(-2-k)\zeta(3-k)\zeta(3+k)\zeta(8+k)}{k!\zeta(6+2k)}\Big(-\frac{\pi}{\tau_2}\Big)^k.\non \\ \eea

Here $k=1,2$ and $3$ contributes to the infinite sum, leading to
\be F_0= \frac{35\zeta(9)}{192\pi^3\tau_2^3} -\frac{45\zeta(5)^2}{64\pi^4\tau_2^4}+\frac{75075\zeta(11)}{88448\pi^5\tau_2^5}\ee
where the $k=2$ term needs regularization. Thus using the expression for the zero mode for $C_{2,2,2}$ in \C{asymp222}, we get that 
\bea \label{asymp}C_{1,2,3} = \frac{473\zeta(12)\tau_2^6}{691\pi^6} + \frac{\zeta(5)\pi\tau_2}{630}+ \frac{\zeta(7)}{144\pi\tau_2} +\frac{7\zeta(9)}{64\pi^3\tau_2^3} -\frac{17\zeta(5)^2}{64\pi^4\tau_2^4} +\frac{99\zeta(11)}{256\pi^5\tau_2^5} \eea
in perfect agreement with the expression in~\cite{DHoker:2016quv}.

\section{The analysis for a linear combination of modular graphs with eight links}

\subsection{The absolutely convergent Poincar\'e series}

As a final example, we consider the linear combination of modular graphs with eight links given by
\be \label{spec}6 C_{1,3,4} + 3 C_{2,2,4} + 5 C_{2,3,3}\ee
which satisfies the eigenvalue equation~\cite{DHoker:2015gmr}
\be \Big(\Delta -2\Big) \Big(6 C_{1,3,4} + 3 C_{2,2,4} + 5 C_{2,3,3}\Big) = 153 E_8 - 9 E_4^2.\ee
We rewrite this as
\be \Big(\Delta -2\Big)\Psi = \frac{65637}{3617} E_8 - 9 E_4^2\ee
where
\be \Psi = 6 C_{1,3,4} + 3 C_{2,2,4} + 5 C_{2,3,3} -\frac{27098}{10851}E_8.\ee
Thus the Poincar\'e series for $\Psi$ defined by
\be \Psi = \sum_{\g \in \G_\infty \backslash SL(2,\mathbb{Z})} \Phi (\g(\tau))\ee
has the seed function which satisfies the equation
\be (\Delta -2) \Phi = \frac{\pi^8}{2480625} \tau_2^8 - \frac{\pi^4}{525}\tau_2^4 E_4\ee
the right hand side of which is $O(\tau_2)$ for large $\tau_2$.
 
Hence defining the Fourier series of the seed function as
\be \Phi (\tau) = \Phi_{0} (\tau_2) + \sum_{n \neq 0} \Phi_{n}(\tau_2) e^{2\pi i n \tau_1}.\ee
we see that the zero mode $\Phi_{0} (\tau_2)$ satisfies
\be \Big(\tau_2^2\frac{d^2}{d\tau_2^2} -2\Big) \Phi_{0} = -\frac{\pi\zeta(7)}{840}\tau_2,\ee
which is solved by
\be \Phi_{0} (\tau_2)= \frac{\pi \zeta(7)}{1680}\tau_2.\ee
The non--zero mode $\Phi_{1,n} (n \neq 0)$ satisfies the differential equation
\bea \Big(\tau_2^2\frac{d^2}{d\tau_2^2} -2 - 4\pi^2 n^2\tau_2^2\Big)\Phi_{1,n} = -\frac{\pi^4\tau_2^4 \s_7(n)}{1575 n^4}e^{-2\pi \vert n \vert \tau_2 } \Big(1+\frac{3}{\pi \vert n \vert \tau_2}+\frac{15}{4\pi^2n^2\tau_2^2} + \frac{15}{8\pi^3\vert n \vert^3\tau_2^3}\Big).\non \\ \eea
As before, expressing
\be \Phi_{n} = \Phi_{n}^h +\Phi_{n}^p,\ee
we see that the solution to the homogeneous equation is given by
\be \Phi_{n}^h = a_{n} \sqrt{\tau_2} K_{3/2}(2\pi \vert n \vert\tau_2)\ee
where $a_n$ is an arbitrary constant.
Now the solution to the particular equation is given by
\bea \Phi_{n}^p = \frac{\s_7 (n)}{302400n^8}\Big( \frac{105}{2\pi \vert n \vert \tau_2} +105 + 180\pi \vert n \vert \tau_2 + 80\pi^2 n^2 \tau_2^2 + 16 \pi^3\vert n \vert^3\tau_2^3 \Big) e^{-2\pi \vert n \vert \tau_2},\eea
and thus the boundary condition for small $\tau_2$ demanding the cancellation of the $\tau_2^{-1}$ term yields
\be a_{n} = -\frac{\s_7 (n)}{1440 \vert n \vert^{15/2}},\ee
leading to the expression
\be \Phi_{n} (\tau_2) = \frac{\pi\s_7 (n)\tau_2}{75600 \vert n\vert^7} \Big(45+ 20\pi \vert n \vert \tau_2 + 4\pi^2 n^2\tau_2^2\Big)e^{-2\pi \vert n \vert \tau_2}\ee
for the non--zero mode.

Thus we have that
\be  6C_{1,3,4} + 3C_{2,2,4}+5C_{2,3,3} - \frac{27098}{10851}E_6= \sum_{\g \in \G_\infty \backslash SL(2,\mathbb{Z})} \L (\g(\tau)),\ee
where the seed function is given by
\be \label{l}\Lambda (\tau) =  \frac{\pi\zeta(7)}{1680}\tau_2 + \frac{\pi \tau_2}{75600} \sum_{n=1}^\infty \frac{\s_7(n)}{n^7} \Big(45+ 20\pi n \tau_2+ 4\pi^2 n^2\tau_2^2\Big)\Big(e^{2\pi i n \tau}+ e^{-2\pi i n \bar\tau}\Big).\ee
To remedy the lack of absolute convergence due to the presence of the term linear in $\tau_2$ in $\Lambda (\tau)$, as in the earlier cases we consider
\be E_4^2 -\frac{7293}{3617}E_8 = \sum_{\g \in \G_\infty \backslash SL(2,\mathbb{Z})} \Upsilon_4 (\g(\tau)),\ee
with the seed function
\be \label{up4}\Upsilon_4 (\tau) = \Lambda_8 (\tau) + \frac{2\zeta(8)\tau_2^4}{3\pi^4} \sum_{n=1}^\infty \frac{\s_7(n)}{n^4} \Big(1+\frac{3}{\pi n \tau_2} +\frac{15}{4\pi^2n^2\tau_2^2}\Big)\Big(e^{2\pi i n \tau}+ e^{-2\pi i n \overline\tau}\Big).\ee
In \C{up4} we have that
\be \Lambda_8 (\tau)= \frac{\pi\zeta(7)}{7560}\tau_2 + \frac{\pi\tau_2}{7560}\sum_{n=1}^\infty \frac{\s_7(n)}{n^7} \Big(e^{2\pi i n \tau}+ e^{-2\pi i n \overline\tau}\Big) \ee
which is exactly proportional to the problematic term in \C{l} which is an obstruction to the absolute convergence of the Poincare series.
 
This yields that
\be 6C_{1,3,4} + 3C_{2,2,4} + 5 C_{2,3,3} +\frac{142715}{21702}E_8-\frac{9}{2} E_4^2= \sum_{\g \in \G_\infty \backslash SL(2,\mathbb{Z})}  \Omega_{8} (\g(\tau)),\ee
where the seed function is
\bea \label{8}\Omega_8 (\tau) &=& -\frac{\zeta(8)\tau_2^4}{\pi^4}\sum_{n=1}^\infty \frac{\s_7(n)}{n^4} \Big(3+\frac{17}{2\pi n \tau_2}+\frac{35}{4\pi^2 n^2\tau_2^2}\Big)\Big(e^{2\pi i n \tau}+ e^{-2\pi i n \overline\tau}\Big)\eea
leading to an absolutely convergent Poincar\'e series.

\subsection{Zero mode of the Fourier series}

We now consider the contribution to the zero mode of the Fourier series that follows from the above analysis. For
\be F = 6C_{1,3,4} + 3C_{2,2,4} + 5 C_{2,3,3} +\frac{142715}{21702}E_8-\frac{9}{2} E_4^2\ee
in \C{Fourier}, from \C{8} and \C{Seed}
we thus have that
\be\label{three} \Omega_0 =0,\quad \Omega_m = -\frac{\zeta(8)\tau_2^4\s_7(m)}{m^4\pi^4}\Big(3+\frac{17}{2\pi \vert m \vert \tau_2}+\frac{35}{4\pi^2 m^2\tau_2^2}\Big)e^{-2\pi\vert m \vert \tau_2}~(m \neq 0).\ee
The contributions to $F_0$ from the three terms in \C{three} that are $O(\tau_2^0)$, $O(1/\tau_2)$ and $O(1/\tau_2^2)$ in the parentheses are given by  
\bea F_0^{(1)} &=& -\frac{\zeta(8)}{64\pi^3\tau_2^3} \sum_{k=0}^\infty \frac{(k+6)(k+5)(k+4)}{k!\zeta(2k+8)} \Big(-\frac{\pi}{\tau_2}\Big)^k \sum_{m>0}\frac{\s_7(m)\s_{-7-2k}(m)}{m^{4-k}} \non \\ &=&-\frac{1}{64\pi^3\tau_2^3}\sum_{k=0}^\infty \frac{(k+6)(k+5)(k+4)\zeta(-3-k)\zeta(4-k)\zeta(4+k)\zeta(11+k)}{k!\zeta(8+2k)}\Big(-\frac{\pi}{\tau_2}\Big)^k,\non \\  F_0^{(2)} &=& -\frac{17\zeta(8)}{32\pi^4\tau_2^2} \sum_{k=0}^\infty \frac{(k+4)(k+3)}{k!\zeta(2k+6)} \Big(-\frac{\pi}{\tau_2}\Big)^k \sum_{m>0}\frac{\s_7(m)\s_{-5-2k}(m)}{m^{5-k}} \non \\ &=&-\frac{17}{32\pi^4\tau_2^2}\sum_{k=0}^\infty \frac{(k+4)(k+3)\zeta(-2-k)\zeta(5-k)\zeta(3+k)\zeta(10+k)}{k!\zeta(6+2k)}\Big(-\frac{\pi}{\tau_2}\Big)^k,\non \\ F_0^{(3)} &=& -\frac{35\zeta(8)}{8\pi^5\tau_2} \sum_{k=0}^\infty \frac{(k+2)}{k!\zeta(2k+4)} \Big(-\frac{\pi}{\tau_2}\Big)^k \sum_{m>0}\frac{\s_7(m)\s_{-3-2k}(m)}{m^{6-k}} \non \\ &=&-\frac{35}{8\pi^5\tau_2}\sum_{k=0}^\infty \frac{(k+2)\zeta(-1-k)\zeta(6-k)\zeta(2+k)\zeta(9+k)}{k!\zeta(4+2k)}\Big(-\frac{\pi}{\tau_2}\Big)^k\non \\ \eea
respectively.

Adding the three contributions, we get that
\bea \label{addf}F_0 = \sum_{i=1}^3 F_0^{(i)} = \frac{5\zeta(9)}{432\pi\tau_2} + \frac{1}{64\tau_2^6} \sum_{k=0}^\infty \frac{g(k)\zeta(-6-k)\zeta(1-k)\zeta(7+k)\zeta(14+k)}{\zeta(14+2k)}\Big(-\frac{\pi}{\tau_2}\Big)^k,\non \\ \eea
where the factor $g(k)$ is given by
\be g(k) = \frac{(k+1)(k+3)(k+7)(k+10)(k+12)}{(k+5)!}.\ee
In \C{addf}, only $k=0$ and 1 contributes to the sum leading to
\be F_0= \frac{5\zeta(9)}{432\pi\tau_2}-\frac{945\zeta(7)^2}{512\pi^6\tau_2^6}+\frac{7300293\zeta(15)}{1851904\pi^7\tau_2^7}\ee
where the $k=0$ term requires regularization.

Thus we get that
\bea 6C_{1,3,4} + 3C_{2,2,4} + 5 C_{2,3,3} = \frac{54196\zeta(16)\tau_2^8}{10851\pi^8} +\frac{\zeta(7)\pi\tau_2}{840} +\frac{5\zeta(9)}{432\pi\tau_2}-\frac{45\zeta(7)^2}{512\pi^6\tau_2^6} +\frac{2431\zeta(15)}{2048\pi^7\tau_2^7}.\non \\ \eea
We have checked this exactly matches the expression one gets by directly analyzing the eigenvalue equation this combination of modular graphs satisfies, as well as with the structure one obtains from~\cite{DHoker:2019txf}. 

For the various examples, the seed functions we have obtained for the Poincar\'e series of the modular graphs are with respect to $\G_\infty \backslash SL(2,\mathbb{Z})$, which is the same coset with respect to which the seed functions for the Poincar\'e series for arbitrary two loop graphs have been obtained in~\cite{DHoker:2019txf}. It would be interesting to understand relations between them.       

\section{Discussion}

In this paper, we have looked at several examples of (linear combinations of) modular graphs where we have calculated the power behaved terms in the zero mode of their Fourier expansion, starting from an expression for the absolutely convergent Poincar\'e series involving them. There are several issues we would like to mention:

(i) We have only looked at graphs with an even number of links, which satisfy eigenvalue equations with source terms involving $E_{2s}$ and $E_s^2$. What about the other cases? For example, one can consider graphs with an odd number of links like $C_{1,1,3}$ which satisfies the eigenvalue equation
\be \Big(\Delta -6\Big) \Big(C_{1,1,3} +\frac{\zeta(5)}{60} \Big) = \frac{86}{5} E_5 - 4 E_2 E_3,\ee     
or graphs with an even number of links like $C_{1,1,4}$ which satisfies the eigenvalue equation
\be \Big(\Delta -12\Big) \Big(6 C_{1,1,4} + C_{2,2,2}\Big) = 120 E_6 +12 E_3^2 - 36E_2 E_4.\ee
In either case, we have performed the analysis and found that it yields the wrong coefficient of only one term in the zero mode of the Fourier series\footnote{For $C_{1,1,3}$ this has been observed before in~\cite{Ahlen:2018wng}.}, even though we use an absolutely convergent Poincar\'e series in \C{Fourier1}. This seems to be the issue whenever the eigenvalue equation contains a source term that is a product of modular graphs that are not identical (for example, $E_{s_1} E_{s_2}$ for $s_1 \neq s_2$ in the examples just mentioned).

In fact, it is for this reason we have chosen the linear combination of graphs in \C{spec} with eight links for our analysis, which has only $E_8$ and $E_4^2$ as the source terms in its eigenvalue equation, even though each graph has source terms in its eigenvalue equation that include other modular graphs as well.    

(ii) This issue naturally raises the question--why did our analysis work for the examples we considered where the source terms only involve $E_{2s}$ and $E_s^2$? While this is perhaps somewhat surprising, our calculations do not lead to a good understanding of why it works. In fact, these issues have been considered from a somewhat different point of view in~\cite{Dorigoni:2019yoq}. As we mentioned after \C{changesum}, in \C{changesum} we have interchanged two infinite sums which is strictly not allowed (we have followed the same strategy to perform similar sums in the other examples as well) due to issues of convergence. In~\cite{Dorigoni:2019yoq}, contributions of this type have been analyzed using results in~\cite{ZagierMellin}, which lead to an extra contribution referred to as the Riemann term. Their analysis involved regularization, which arises from modifying the original eigenvalue equation infinitesimally in an $SL(2,\mathbb{Z})$ invariant way. While this leads to a somewhat involved analysis, this yields the correct answer for all the cases. For the cases we have looked at, such additional contributions that arise beyond the ones we have considered must cancel\footnote{I am thankful to the anonymous referee for illuminating comments on this issue.} leading to the simplification. It would be interesting to understand these issues in detail for general modular graphs.             
 
(iii) We only focussed on the power behaved terms in the Fourier expansion of the zero modes, and neglected the contributions that are exponentially suppressed in the large $\tau_2$ limit which did not arise in our analysis. In fact, in the final expression for $F_0$ only a few terms contributed to the infinite sum. In~\cite{Dorigoni:2019yoq} it has been suggested that the exponentially suppressed terms can be obtained precisely from these terms that vanish, based on certain assumptions and using the regularization mentioned in (ii). Once again, it would be interesting to understand these issues for our analysis.

(iv) Thus we see that the results we have obtained for several examples yield the correct power behaved terms in the zero mode of the Fourier series, which is really an experimental observation. Our results suggest an obvious generalization which we briefly mention. Consider a linear combination of modular graphs $\Psi_{2s}$ with $2s$ links which satisfies the eigenvalue equation\footnote{If the source terms in the eigenvalue equation involve only $E_{2s}$ and $E_s^2$, we can always bring it into this form by rescaling $\Psi_{2s}$ and then shifting $E_{2s}$ with an appropriate coefficient into the definition of $\Psi_{2s}$ on using \C{Eisenval}.}
\be \Big(\Delta - s'(s'-1)\Big) \Psi_{2s} = E_{2s} - \frac{\zeta(4s)}{2\zeta(2s)^2} E_s^2,\ee  
where $s'$ is an integer greater than 1. Thus defining the Poincar\'e series
\be \Psi_{2s} (\tau) = \sum_{\g \in \G_\infty \backslash SL(2,\mathbb{Z})} \Phi_{2s}(\g(\tau)), \ee
we see that the seed function satisfies the equation
\be \Big(\Delta - s'(s'-1)\Big) \Phi_{2s} = \frac{2\zeta(4s)}{\pi^{2s}}\tau_2^{2s} - \frac{\zeta(4s)\tau_2^s}{\zeta(2s)\pi^{s}}E_s,\ee
where the right hand side is $O(\tau_2)$ for large $\tau_2$. Then one can proceed as we did with the choice of boundary conditions for the seed function to obtain the power behaved terms in the zero mode of the Fourier series. It is an interesting exercise to see if this yields the correct answer, hence circumventing the need for regularization.

\appendix

\section{The $SL(2,\mathbb{Z})$ invariant non--holomorphic Eisenstein series}

The non--holomorphic Eisenstein series $E_s (\tau)$ satisfies the eigenvalue equation
\be \label{Eisenval}\Delta E_s = s(s-1)E_s,\ee
and has the Fourier expansion
\bea \label{Eisen}E_s (\tau) &=& \frac{2\zeta(2s)}{\pi^s}\tau_2^s +  \frac{2\Gamma(s-1/2)}{\pi^{s-1/2}\Gamma(s)}\zeta(2s-1) \tau_2^{1-s}\non \\ &&+ \frac{4\sqrt{\tau_2}}{\Gamma(s)}\sum_{n \neq 0} \frac{\s_{2s-1}(n )}{\vert n\vert^{s-1/2}}K_{s-1/2}(2\pi \vert n\vert \tau_2) e^{2\pi i n\tau_1},\eea
where the divisor function $\s_m (n)$ is defined by
\be \s_m (n) = \sum_{d|n, d >0} d^m,\ee
where the sum is over the positive divisors of $n$.

The Poincar\'e series representation for the Eisenstein series $E_s (\tau)$ which is absolutely convergent for ${\rm Re} s > 1$ is given by
\be \label{P}E_s (\tau) = \frac{2\zeta(2s)}{\pi^s}\sum_{\g \in \G_\infty \backslash SL(2,\mathbb{Z})} ({\rm Im}\g(\tau))^s,\ee
where
\be \g(\tau) = \frac{a\tau+b}{c\tau+d}\ee
is an $SL(2,\mathbb{Z})$ transformation, under the identification by
\be \G_\infty = \pm \begin{pmatrix}1 & n \\ 0 & 1\end{pmatrix},\ee
for $n \in \mathbb{Z}$ which stabilizes the cusp at $\tau_2\rightarrow \infty$. 

\section{The Fourier series from the Poincar\'e series}

Suppose $F(\tau)$, which is modular invariant, has the Fourier expansion
\be \label{Fourier}F(\tau) = \sum_{n \in \mathbb{Z}} F_n (\tau_2) e^{2\pi i n \tau_1}.\ee
Also suppose $F(\tau)$ is given by the absolutely convergent Poincar\'e series
\be F(\tau) = \sum_{\g \in \G_\infty \backslash SL(2,\mathbb{Z})} \Omega (\g(\tau)),\ee
where the seed function $\Omega(\tau)$ has the Fourier expansion
\be \label{Seed}\Omega(\tau) = \sum_{n\in \mathbb{Z}} \Omega_n (\tau_2)e^{2\pi i n \tau_1}.\ee
Then the Fourier modes $F_n (\tau_2)$ of $F(\tau)$ can be obtained from the Fourier modes $\Omega_n (\tau_2)$ of the seed function $\Omega(\tau)$. 

Focussing only on $F_0(\tau_2)$, we have that (see\cite{Iwaniec,Fleig:2015vky}, for example)
\bea \label{Fourier1}F_0(\tau_2) = \Omega_0 (\tau_2) + \tau_2 \sum_{n > 0} \sum_{m \in \mathbb{Z}} S(m,0;n) \int_{-\infty}^\infty dx \Omega_m \Big(\frac{1}{n^2\tau_2(1+x^2)}\Big)e^{-2\pi i mx/[n^2\tau_2(1+x^2)]},\non \\ \eea 
where $S(m,0;n)$ is obtained from the the Kloosterman sum
\be \label{Ksum}S(m,n;p) = \sum' e^{2\pi i(am+bn)/p}.\ee
In \C{Ksum}, the restricted sum is over the integers
\be 0 < a < p, \quad (a,p)=1, \quad ab= 1~({\rm mod}p).\ee

In our analysis, we need the expression for the Kloosterman--Selberg zeta function given by 
\be \label{Fourier2}\sum_{n>0} \frac{S(\pm m,0;n)}{n^{2s}} = \frac{\s_{1-2s}(m)}{\zeta(2s)}, ~(m \neq 0).\ee


\providecommand{\href}[2]{#2}\begingroup\raggedright\endgroup

\end{document}